\documentclass[12pt]{article}
\usepackage{doc}
\usepackage{epsf}
\usepackage[T1]{fontenc}
\usepackage[finnish,english]{babel}
\usepackage[dvips]{graphicx}
\usepackage{amsmath}
\usepackage{amssymb}
\usepackage{verbatim}

\title{Dynamics and Kinetic Roughening of Interfaces in Two-Dimensional Forced Wetting}
\author{T. Laurila$^1$, C. Tong$^1$ I. Huopaniemi$^1$, S. Majaniemi$^{1,2}$ 
and T. Ala-Nissila$^{1,3}$\\
$^1$Laboratory of Physics, P.O. Box 1100,\\
Helsinki University of Technology, FIN--02015 HUT,\\
Espoo, Finland \\
$^2$Department of Physics, McGill University, 3600 rue University,\\
Montreal, QC, Canada H3A 2 \\
$^3$Department of Physics, P.O. Box 1843, Brown University, \\
Providence, RI 02912--1843, U.S.A. }

\begin{document}

\date{April 7, 2005}

\maketitle

\begin{abstract}
We consider the dynamics and kinetic roughening of wetting fronts in
the case of forced wetting driven by a constant mass flux into a 2D
disordered medium. We employ a coarse-grained phase field model with
local conservation of density, which has been developed earlier for
spontaneous imbibition driven by a capillary forces. The forced flow
creates interfaces that propagate at a constant average velocity. We
first derive a linearized equation of motion for the interface fluctuations
using projection methods. From this we extract a time-independent
crossover length $\xi_\times$, which separates two regimes of dissipative
behavior and governs the kinetic roughening of the interfaces by
giving an upper cutoff for the extent of the
fluctuations. By numerically integrating the phase field model, we
find that the interfaces are superrough with a roughness exponent of
$\chi = 1.35 \pm 0.05$, a growth exponent of $\beta = 0.50 \pm
0.02$, and $\xi_\times \sim v^{-1/2}$ as a function of the velocity. 
These results are in good agreement with recent experiments
on Hele-Shaw cells. We also make a direct numerical comparison
between the solutions of the full phase field model and the
corresponding linearized interface equation. Good agreement is found
in spatial correlations, while the temporal correlations in the two
models are somewhat different.
\end{abstract}

\section{Introduction}

The dynamics and roughening of moving interfaces in a disordered
medium is a subject of intense interest in non-equilibrium
statistical physics \cite{barabasi95}. Examples where such processes
are relevant include thin film deposition \cite{mg00}, slow
combustion fronts in paper \cite{Myllys00}, fluid invasion in
fractals \cite{asikainen02} and porous media
\cite{bg90,sch57,krug97}, and wetting and propagation of contact
lines \cite{mrkr,jr90,ek94}. The understanding of the underlying
physics involved in interface roughening is crucial to the control
and optimization of these processes, with immediately apparent
technological importance. Progress in the theoretical study of
interface dynamics has been made over the last two decades and a
number of theories have been developed \cite{barabasi95,krug97}
which agree with the experimental findings in some cases. Much of
the theoretical work has been based on analyzing (spatially) local
interface equations, such as the celebrated Kardar-Paris-Zhang (KPZ)
equation \cite{kpz}, where the physics is governed by a (nonlinear) partial
differential equation which couples the interface {\it locally} with
itself and the quenched randomness. However, in many cases such a
description is not possible \cite{asikainen02,dubeprl}.

A particularly important class of problems in the field of kinetic
roughening where local theories cannot provide a complete
description are those involving fluid invasion in porous media,
which are often experimentally studied using Hele-Shaw cells
\cite{SOH1,soh,Hernandez-machado01,GMH02} or even paper as the wetting medium
\cite{bb92,hs95,amaral94,KHO96}. The reason for this is that if the
transport of liquid to the advancing wetting front from the
reservoir is neglected as in local theories, slowing down of the
front in spontaneous imbibition of water in paper cannot be
explained by local theories unless the liquid conservation law is
included in some artificial way. To properly describe the dynamics
of wetting fronts in random medium is a challenging task, and there
are several recent theoretical attempts to this end
\cite{Hernandez-machado01,pc03,ganesan98}. In particular, in Refs.
\cite{dubeprl,dube00a,dube01,AME04} Dube {\it et al.} developed a
phase-field model explicitly addressing the issue of liquid
conservation in the wetting of a random medium. This is achieved by
a generalized Cahn-Hilliard equation with suitable boundary
condition which couples the system to the reservoir. A variant of
the sharp interface projection method \cite{kawasaki82,elder01} was used to
analytically obtain a non-local interface equation for the case of
spontaneous imbibition, and from it a new time dependent length scale governing the
kinetic roughening $\xi_\times \sim t^{1/4}$, was extracted. In
Refs. \cite{dubeprl,dube01} the kinetic roughening of 2D wetting
fronts was also analyzed, and estimates for the corresponding
scaling exponents were obtained. Furthermore, in a recent
comprehensive review paper \cite{alava04} the case of forced wetting
in 2D was briefly discussed, and in Ref. \cite{dube05} forced
wetting in a 3D model of paper was also considered.

In the present work our aim is to carry out a comprehensive analysis
of wetting fronts in 2D in the case of forced wetting. To this end,
we use the phase-field model of Ref. \cite{dubeprl} with boundary
conditions corresponding to a constant mass flux at the reservoir
boundary. This makes the wetting fronts move at a constant average
velocity, in contrast to the Washburn law for spontaneous wetting.
We first expand on the standard projection method to make it usable
under constant flow boundary conditions, and derive the linearized
interface equations corresponding to the forced case. Analysis of
these equations reveals a crossover length scale $\xi_\times$, which
is time-independent in contrast to the spontaneous wetting case
%where $\xi_\times \propto t^{1/4}$ 
\cite{dubeprl}, and depends on the interface velocity as
$\xi_\times \propto v^{-1/2}$. It separates two
regimes of dissipative behavior and governs the kinetic roughening
of the interfaces by giving an upper cutoff for the correlation
length of the interface fluctuations. By numerically integrating the
phase field model, we find that the interfaces are superrough with a
roughness exponent of $\chi = 1.35 \pm 0.05$, and a growth exponent
of $\beta = 0.50 \pm 0.02$. These results are in good agreement with
recent experiments on Hele-Shaw cells \cite{soh}. We also make a direct
numerical comparison between the solutions of the full phase field
model and the corresponding linearized interface equation. Good
agreement is found in spatial correlations, while the temporal
correlations in the two models are somewhat different.

This paper proceeds as follows: In Section \ref{model} we define the
phase field model, and present results from the projection and
linearization procedure. We present the resulting interface
equations, as well as discuss how the crossover length scale
$\xi_\times$ emerges. In Section \ref{numerics} we present our
numerical results for the driven interfaces in disordered medium. We
consider both spatial and temporal correlation functions, as well as
compare the numerical results from the linearized interface equation
to the phase field model. Finally, we present our discussion and
conclusions in Section \ref{conclusions}. The Appendices A and B
contain some technical details on the projection and linearization
procedures leading to the interface equations.

\section{Model for Wetting}
\label{model}

\subsection{Definition of the Phase Field Model}

The model describes the dynamics of a liquid invading a disordered
medium at a coarse-grained level. A phase field is used to
describe the ``wet'' and ``dry'' phases, with a free energy
functional such that the dimensionless phase field obtains the
values $\phi=+1$ and $\phi=-1$ at the wet and dry phases,
respectively. Since the phase field is an effective density field
it is locally conserved. Energy cost for an interface is added by
the standard coarse-grained gradient squared term. The interaction
energy between the random medium and the invading liquid is
represented by a quenched random field linearly coupled to the
phase field. This leads to the free energy density \cite{dubeprl}
\begin{equation}
\label{free energy}
\mathcal{F}[\phi(\mathbf{x},t)]=\frac{1}{2}(\nabla\phi(\mathbf{x,t}))^2+V(\phi(\mathbf{x,t}))
-\alpha(\mathbf{x})\phi(\mathbf{x},t),
\end{equation}
where $V$ has two minima at $\phi=-1$ and $\phi=+1$, and $\alpha$ is the
quenched random field. The standard Ginzburg-Landau form is chosen for $V$,
{\it i.e.} $V(\phi)=-\phi^2/2+\phi^4/4$ \cite{dubeprl}, and the quenched
field obeys the relations
\begin{eqnarray}
\label{random field}
\langle\alpha(\mathbf{x})\rangle&=&0\\
\langle\alpha(\mathbf{x})\alpha(\mathbf{x}')\rangle&=&(\Delta \alpha)^2 
\delta(\mathbf{x}-\mathbf{x}').
\end{eqnarray}
The case of positive (negative) $\langle\alpha(\mathbf{x})\rangle$
corresponds to the liquid spontaneously wetting (dewetting) the
medium for the case of no external driving \cite{dubeprl}.

The equation of motion for the conserved phase field is given by
the continuity equation $\partial_t\phi=-\nabla\cdot \mathbf{j}$
and Fick's law $\mathbf{j}=-\nabla\mu$, where
$\mu=\delta\cal{F}/\delta\phi$. The result is
\begin{equation}
\label{pf_eq}
\partial_t \phi({\bf x},t) = \nabla^2 \mu({\bf x},t) = \nabla^2 \left[
- \phi({\bf x},t) + \phi^3({\bf x},t) - \nabla^2 \phi({\bf x},t) -
\alpha({\bf x})\right],
\end{equation}
which is essentially the Cahn-Hilliard equation \cite{cahn58}. Note
that dimensionless units have been set such that the constant
relating the phase field gradient to the free energy and the
mobility in Fick's law are both unity. This fixes the choice of
units in Eq. \eqref{pf_eq}.

In Refs. \cite{dubeprl,dube00a} the case of spontanous,
capillary driven wetting was modeled with Eq. \eqref{pf_eq} using
boundary conditions where the chemical potential was set to a
constant at the liquid reservoir $y=0$. In order to model the
experimental setup of driving the liquid from the reservoir, we
define our phase-field in the half-plane $\{\mathbf{x}\vert y\ge
0\}$, and at the line $y=0$ impose the boundary condition
$\nabla\mu = -F\hat{y}$, where $\hat{y}$ is the unit vector, and
$F$ is a constant (flux) parameter (see Fig. 1(a)). On the top end
of the system we set $\phi(y\rightarrow\infty)=-1$, and use
periodic boundaries in the $x$ direction. Physically this
corresponds to driving the liquid via a constant mass flow,
leading to an interface propagating at a constant average velocity
\cite{alava04}.

The initial condition for the phase field is given by a step
function at some height $H(0)=H(t=0)$,
$\phi(\mathbf{x},t=0)=1-2\Theta(y-H(0))$. $H(0)$ is also a
parameter in our model, but with the gradient boundary condition
here its value is irrelevant, in contrast to the case of
spontaneous imbibition, where $H(0)$ defines the initial average
velocity of the interface \cite{dubeprl}.

\subsection{Linearized interface equation}

The quenched disorder field $\alpha(\mathbf{x})$ will cause an
initially flat interface to kinetically roughen when it propagates
as the liquid invades the medium. A key step in understanding the
physics of this process is writing an equation of motion for the
1D single-valued height variable $H(x,t)$, defined conveniently by
the condition $\phi(x,H(x,t))=0$ as shown in Fig. 1(a). It is not
obvious {\it a priori} that this can be done, since the problem is
inherently non-local due to the conservation law \cite{dubeprl},
but it has been shown in Refs. \cite{dubeprl} that this
can be done using methods discussed in Refs.
\cite{kawasaki82,bray94,langer77}. In Appendix \ref{A} we will
discuss some technical details of the derivation. The main idea is
to use the relevant Green's function of the problem defined by
\begin{equation}
\nabla^2  G(x,y|x',y') = \delta (x-x') \; \delta(y-y'),
\end{equation}
with appropriate boundary conditions. This leads to the
integro-differential equation of motion
\begin{equation}
2\int_{-\infty}^{\infty} dx' \; G (x,H(x,t)|x',H(x',t)) \;
\frac{\partial H(x',t)}{\partial t} = \alpha (x,H(x,t)) + \sigma \kappa
+\Lambda\vert_{y=H(x,t)},
\label{non-local}
\end{equation}
where $\kappa$ is the interface curvature, and $\Lambda$ is the
boundary term, which is non-vanishing for inhomogenous boundary
conditions. Note that Eq.~\eqref{non-local} holds for any geometry
given the appropriate Green's function. For the half plane with
von Neumann boundary condition at $y=0$, the 2D Green's function
is obtained by image charge method as
\begin{equation}
\label{half-plane Green}
G(x,y|x',y') = \frac{1}{4\pi} \ln\left[((x-x')^2 +
(y-y')^2)((x-x')^2+(y+y')^2)\right].
\end{equation}
The next step towards an explicit interface equation is to
linearize in fluctuations around the disorder-free system
solution, $H(x,t)=H_0(t)+h(x,t)$, and transform to Fourier space,
where the equations of different modes of fluctuation are
decoupled. The half plane Green's function defined in Eq.
(\ref{half-plane Green}) is not square integratable, however, and
thus it does not have a Fourier representation. However, we have
found that we can avoid this problem by considering a finite strip
of width $L$ where $0 \le x \le L$, and derive a linearized
equation of motion for the {\it discrete} Fourier modes of the
fluctuations. Then we can take the limit $L \rightarrow \infty$ to
obtain the equation of motion in Fourier space for an interface in
half-plane geometry, and the end result is well-defined. This
procedure is exposed in some detail in Appendix \ref{B}.

The linearization procedure, by construction, gives a separate
equation of motion for the mean interface position $H_0(t)$, which
turns out to be coupled to all the fluctuation Fourier modes
$h_k$, while the evolutions of the fluctuation modes are
independent of each other. The resulting equations are given by
\begin{eqnarray}
\label{ave_int}
\dot{H_0}&=&\frac{F}{2}; \\
\label{fluct_int}
\dot{h}_k \left( 1 + e^{-2 \vert k \vert H_0} \right)&=&
|k| \left( -\dot H_0 \; h_k \left( 1 - e^{-2 \vert k \vert H_0}
\right) - \sigma k^2 h_k + \eta_k (t) \right),
\end{eqnarray}
from which we immediately obtain the expected result that
\begin{equation}
\label{linear}
 H_0(t) = \frac{Ft}{2}.
\end{equation}
This should be contrasted with the Washburn law $H_0(t) \propto
t^{1/2}$ in the case of spontaneous wetting \cite{dubeprl}. It is
interesting to note that the functional form of Eq.
\eqref{fluct_int} for the fluctuations of the interface is similar
to the case of spontaneous wetting in Ref. \cite{dubeprl}, except
for some sign changes in the terms. It can be shown that these
changes are a direct consequence of the differences between the
Green's functions in the two cases, and they lead to significant
differences in the behavior of the fluctuations as will be
discussed below.

The most immediate aspect of these equations is that the
fluctuation equation is non-local in real space. This is to be
expected due to the conservation law \cite{dubeprl}. The locality
of the interface equation in Fourier space owes to the fact that
it has been linearized. The interface configuration couples to the
disorder in a fundamentally non-linear manner, a fact that is
somewhat obscured by the superficially simple form of the disorder
term $\eta_k$, which is defined as
\begin{equation}
 \eta_k (t) \equiv \int dx~e^{-ikx} \alpha(x,H(x,t))
%=\int dx e^{ikx} \frac{1}{2\pi}\int dk_y e^{-ik_yH(x,t)}\int dy e^{ik_yy}\alpha(x,y).
\end{equation}
The moments of $\eta_k$ averaged over disorder realizations thus
couple to the interface configuration realizations, which in turn
are defined by the disorder. This makes the analysis of $\eta_k$ a
formidable task that must be solved self-consistently involving
the interface equation. In more explicit terms, in expressions
such as $\langle \alpha(x,H(x,t))\alpha(x',H(x',t'))\rangle$, the
angular brackets denote averages over different realizations of
$H$, and results can only be obtained numerically.

In Eq. \eqref{fluct_int} there are two terms that dissipate
fluctuations corresponding to different physical effects: the
surface tension term $\sigma k^3 h_k$ and the liquid transport
(conservation law) $|k| \dot H_0 \left( 1 - e^{-2 \vert k \vert
H_0}\right) h_k$. The surface tension dominates when
$k\dot{H_0}\ll \sigma k^3$, leading to a {\it time independent} crossover length
scale between the two terms given by
\begin{equation}
\label{cross}
\xi_\times=2\pi\sqrt{\frac{\sigma}{\dot{H}_0}}=2\pi\sqrt{\frac{2\sigma}{F}}.
\end{equation}
This is in striking contrast to the spontaneous case, where the
corresponding crossover length is {\it time dependent} with
$\xi_\times \propto t^{1/4}$ \cite{dubeprl}. Moreover, in the
dispersion relation of the fluctuations there is an additional
crossover in the transport term obtained by comparing the length
scales $H_0(t)$ and $k$. Namely, for $kH_0(t)\gg 1$ the transport
term is $k\dot{H}_0$, while for $kH_0(t)\ll 1$ it is
$2k^2H_0\dot{H}_0$, leading to a crossover between $\omega\propto
k$ and $\omega\propto k^2$ in the dispersion relation of the
interface fluctuations. Regardless of the magnitude of $kH_0(t)$,
we will show through numerical studies that the crossover length
scale $\xi_\times$ controls the kinetic roughening of the
interface in analogy to the spontaneous imbibition case, in that
it defines an upper cut-off for fluctuations that are increasing
in time and correlated by the surface tension.

\section{Numerical analysis}
\label{numerics}

\begin{figure}[t]

\includegraphics [height = 6cm, width = 6cm]{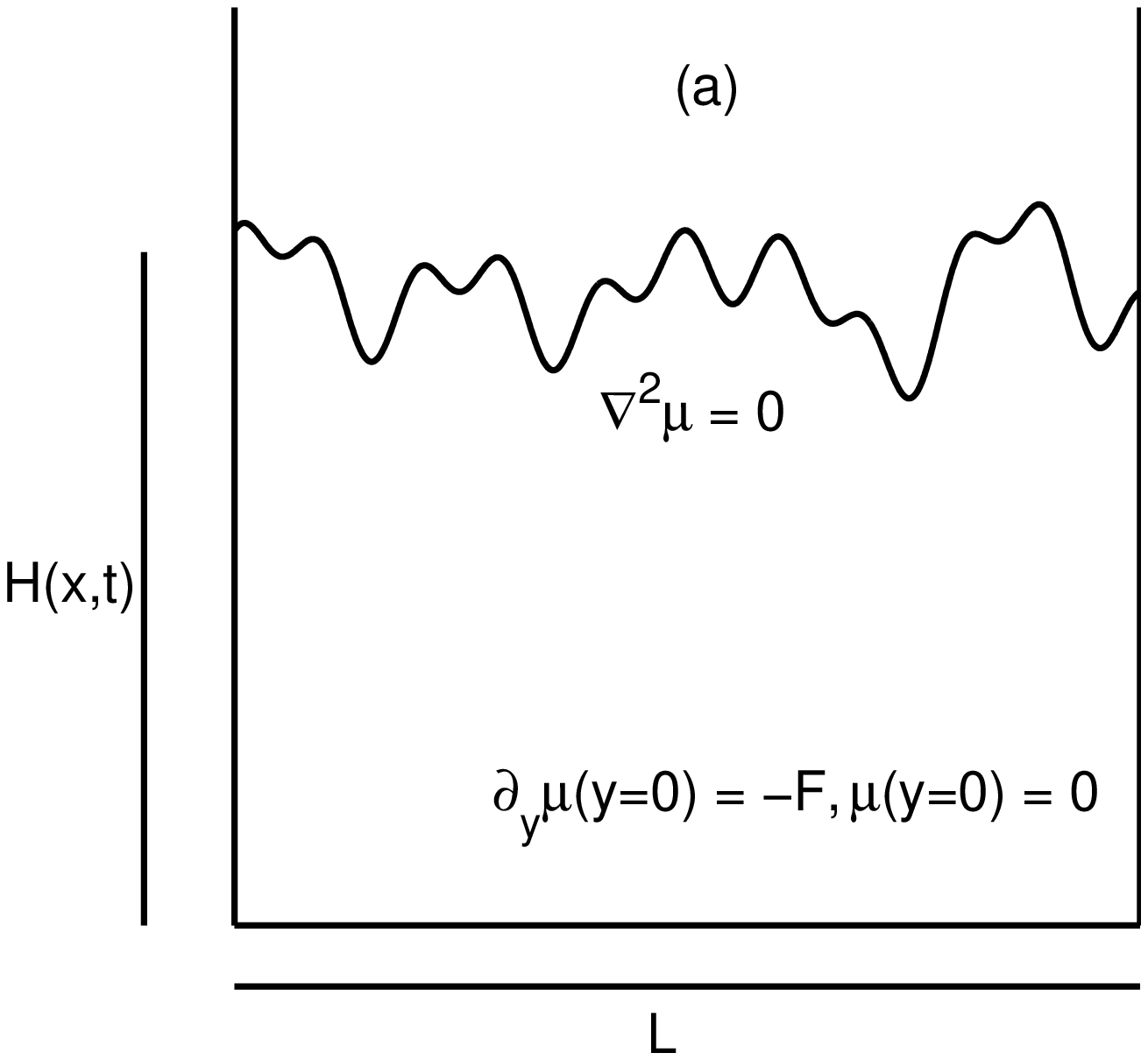}
\includegraphics [height = 6cm, width = 6cm]{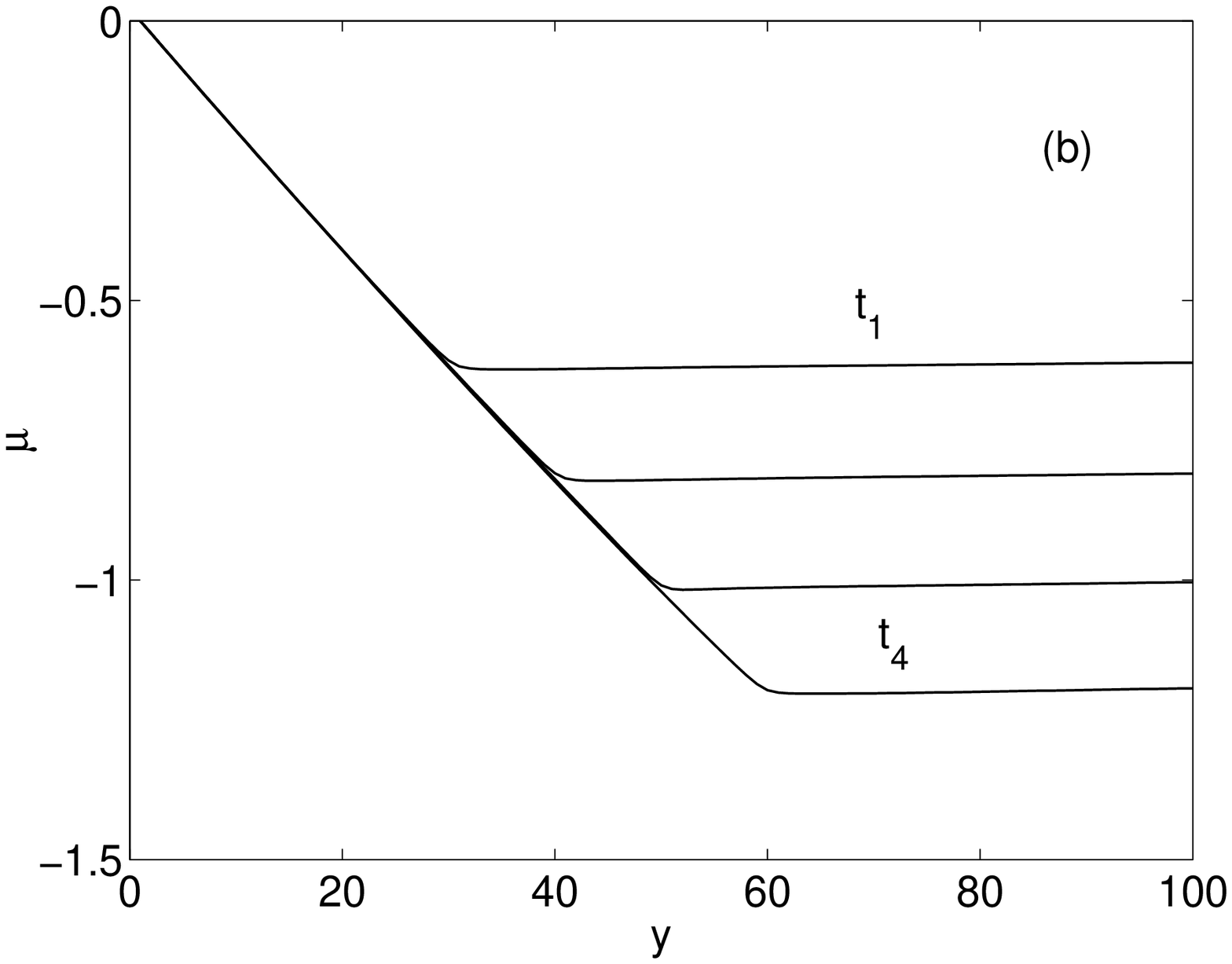}
\includegraphics [height = 6cm, width = 6cm]{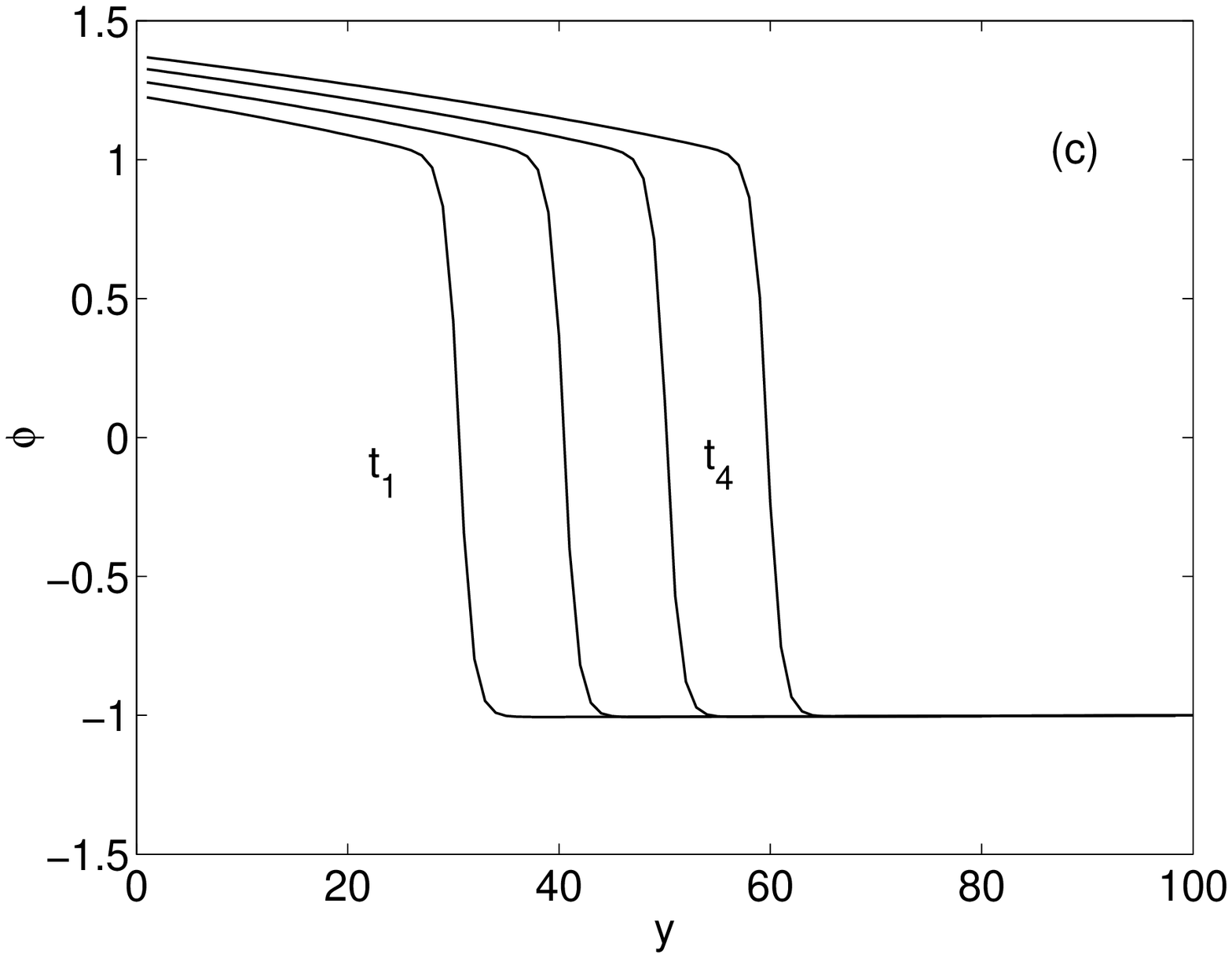}
\includegraphics [height = 6cm, width = 6cm]{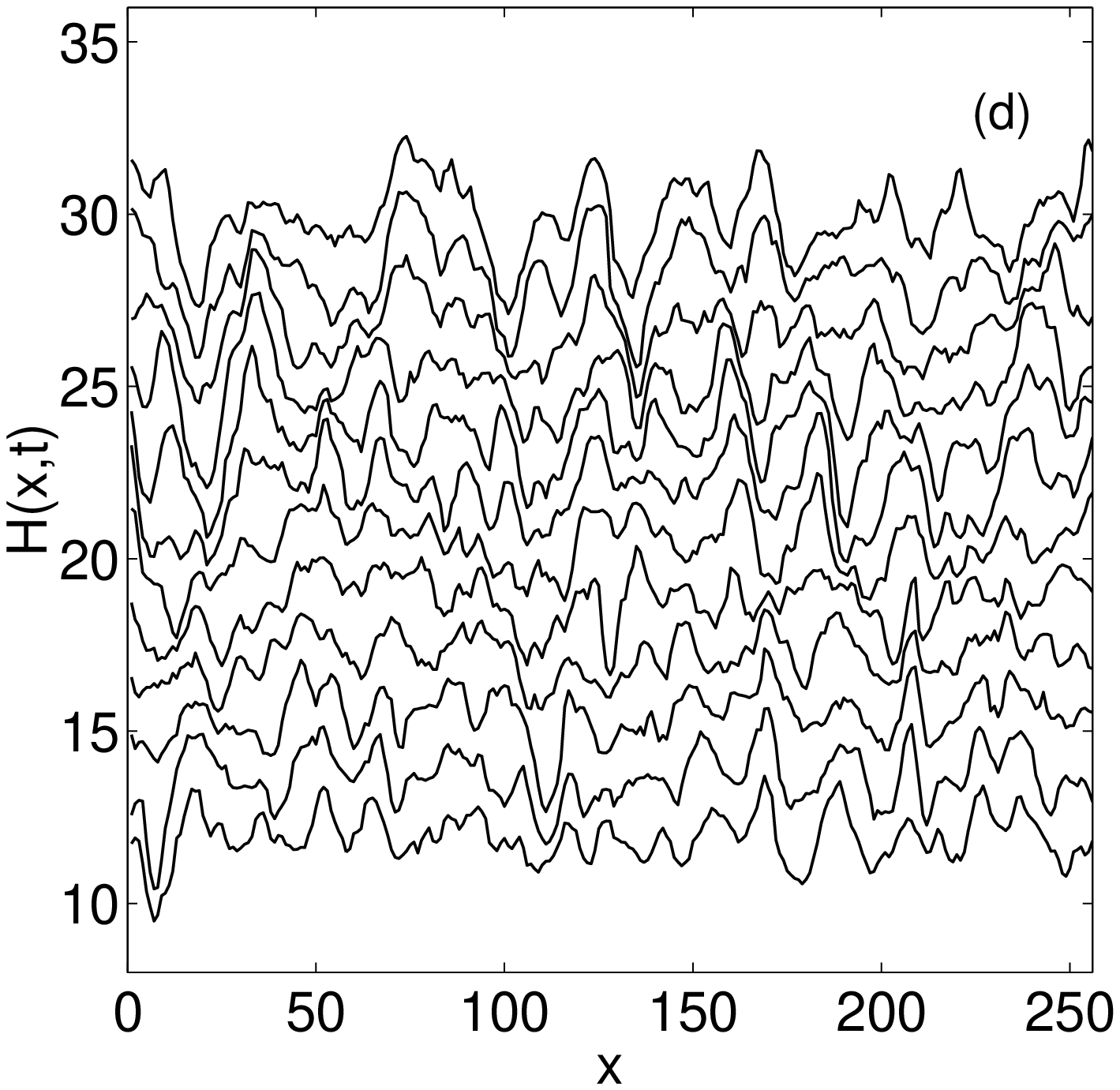}
\caption{
(a) A schematic geometry and setup of the model. The
height of the front is described by a single-valued function
$H(x,t)$, and the driven boundary condition at the reservoir
at $y=0$ is described by a constant gradient of the chemical potential.
(b) The profile of the chemical potential $\mu(x,y)$ along the
$y$ axis at successive time steps $t_1 < t_2 ... < t_4$.
(c) The profile of the density field $\phi(x,y)$ along the
$y$ axis at successive time steps corresponding to (b). Note that due to
the conservation law these profiles have a finite slope in the wet region
of the medium.
(d) A set of typical rough front configurations of a rising interface $H(x,t)$
taken at equal time intervals $ \Delta t = 80$.}
\label{Configuration}
\end{figure}

\begin{figure}[t]
\includegraphics [height = 6cm, width = 6cm]{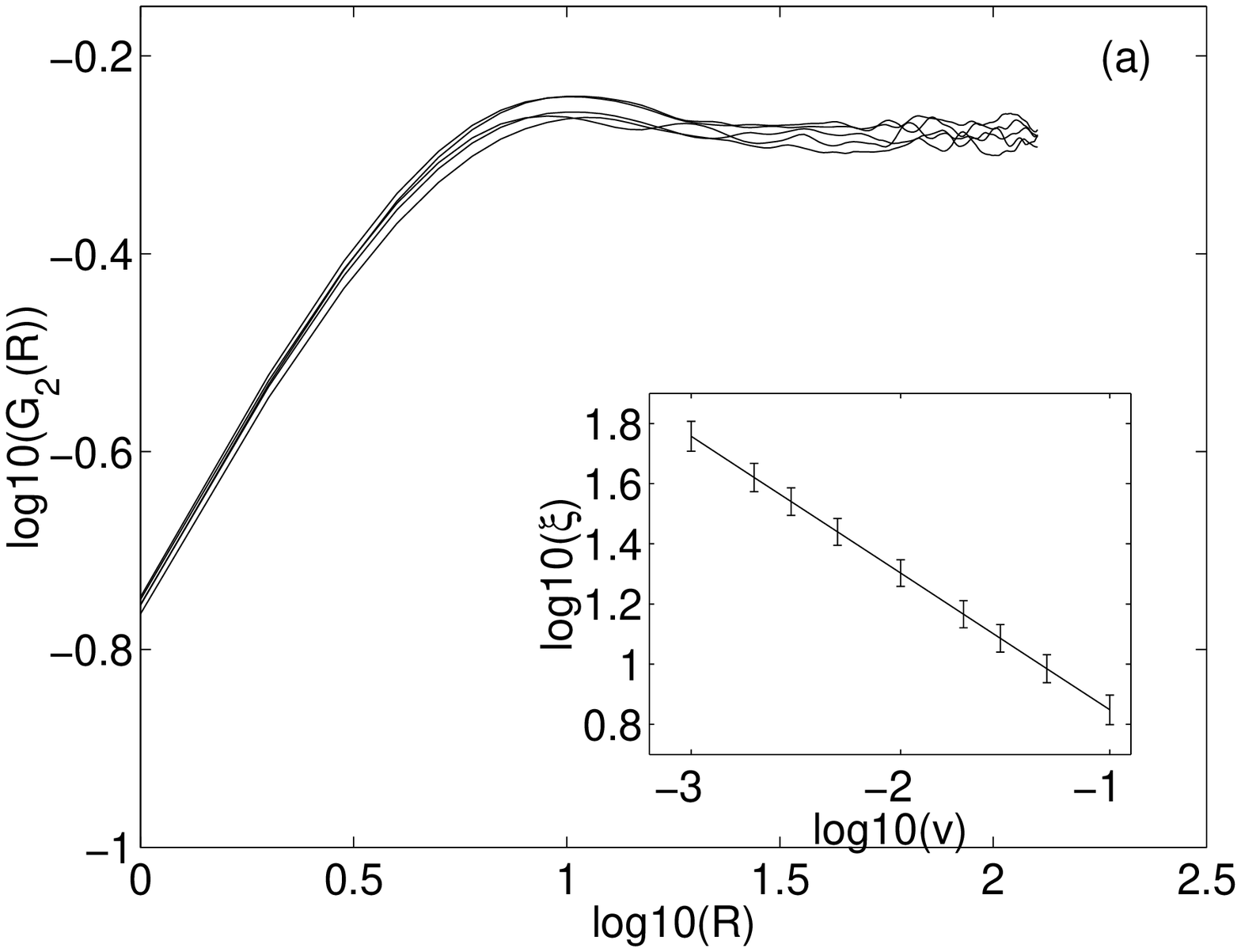}
\includegraphics [height = 6cm, width = 6cm]{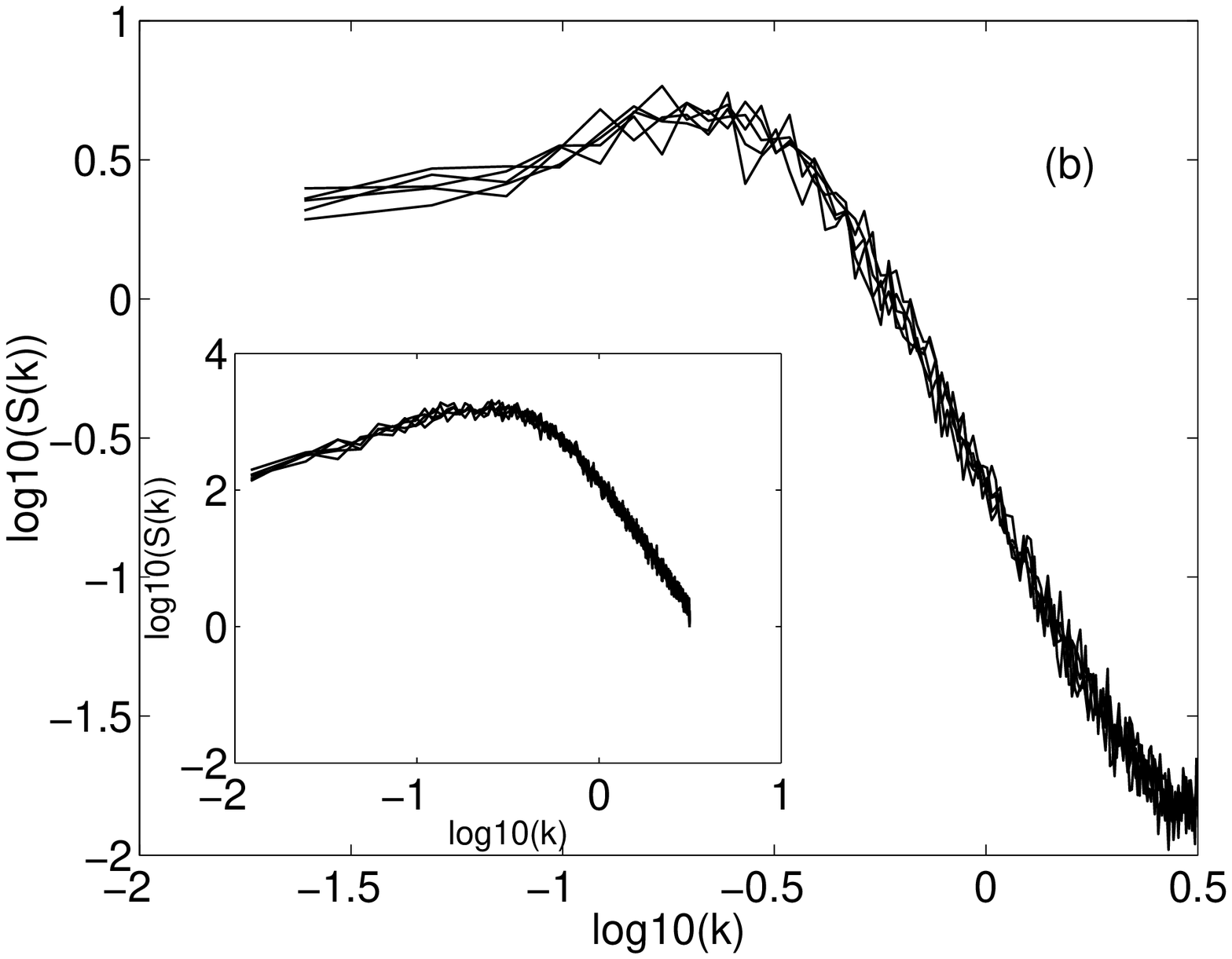}
\caption{(a) Spatial correlation functions $G_2(r,t)$ for a system
of size $L = 256$, with $\Delta \alpha = 0.3$, $v = 0.05$ at
different times. The data are from $t = 10^3$ to $ t = 10^4 $ at
equal time intervals of $ t = 2 \times 10^3$. In the inset, the
crossover lengths $\xi_\times(V)$ obtained from $G_2$ for
different velocities are plotted. (b) The
structure factor $S(k,t)$ plotted against the wave vector $k$ for
the same set of parameters as in (a). In the inset, the structure
factor obtained from the linearized interface equations is plotted
against $k$ for the corresponding set of parameters, except that
$L = 512$.} \label{CorrelationFigure}
\end{figure}

\begin{figure}[t]
\includegraphics [height = 6cm, width = 6cm]{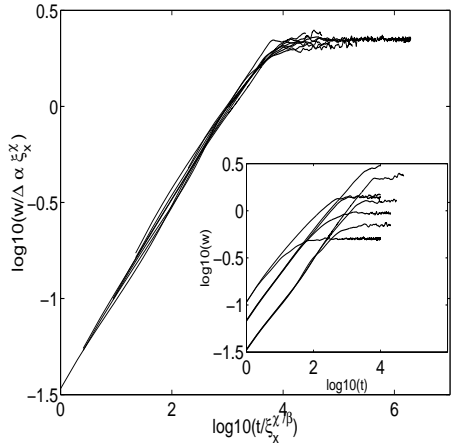}
\caption{Data collapse of the interface width according to the
scaling form of Eq. \eqref{CollapseForm} for different sets of
parameters with $L=256$: (i) $\Delta \alpha = 0.3$, $v= 0.05$ and
$0.01$; (ii) $\Delta \alpha = 0.2$, $v= 0.01, 0.005$, and $0.002$;
(iii)  $\Delta \alpha = 0.1$, $v= 0.005, 0.002$, and $0.001$. The
inset shows the original interface width data. }
\label{DataCollapse}
\end{figure}

\begin{figure}[t]
\includegraphics [height = 6cm, width = 6cm]{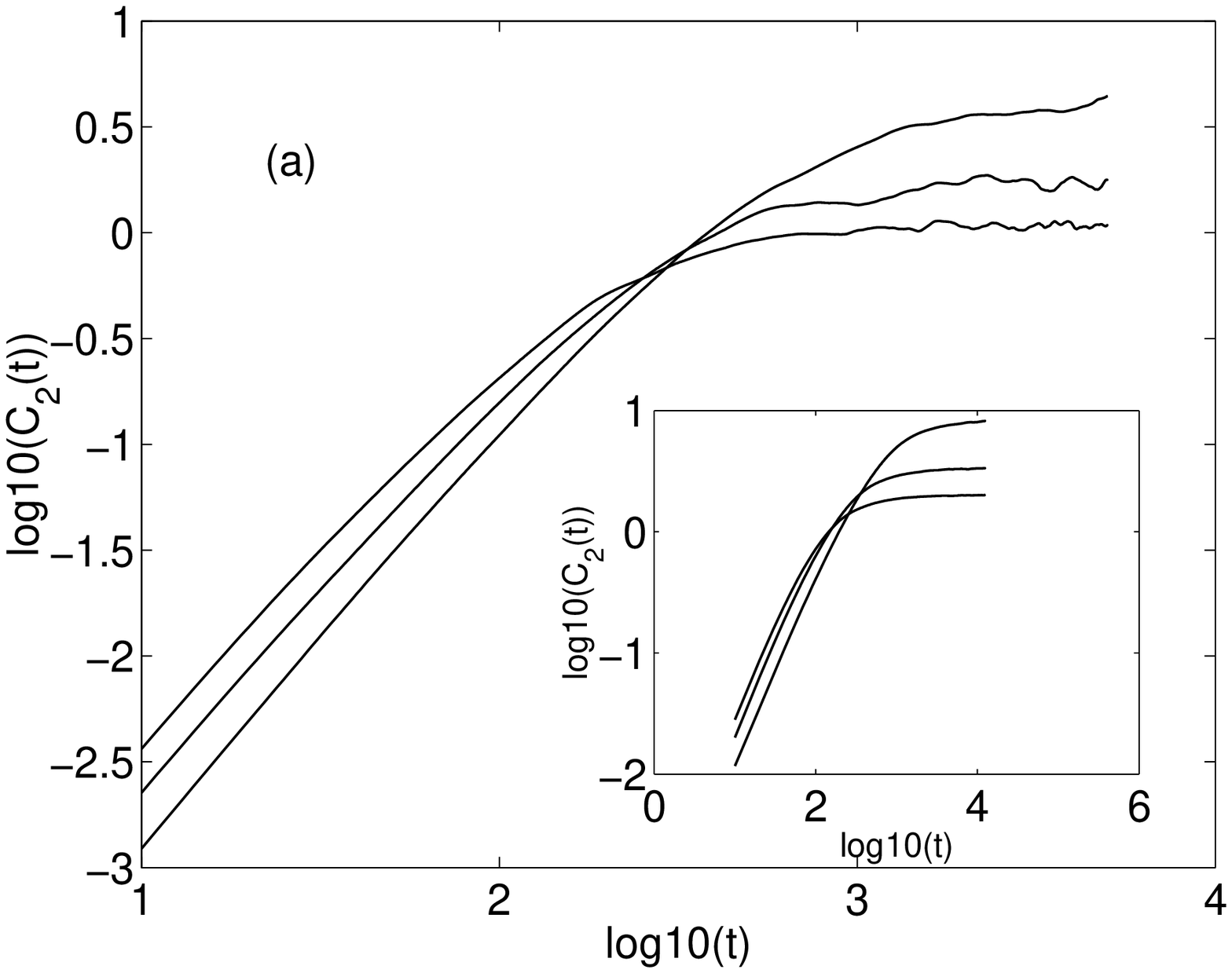}
\includegraphics [height = 6cm, width = 6cm]{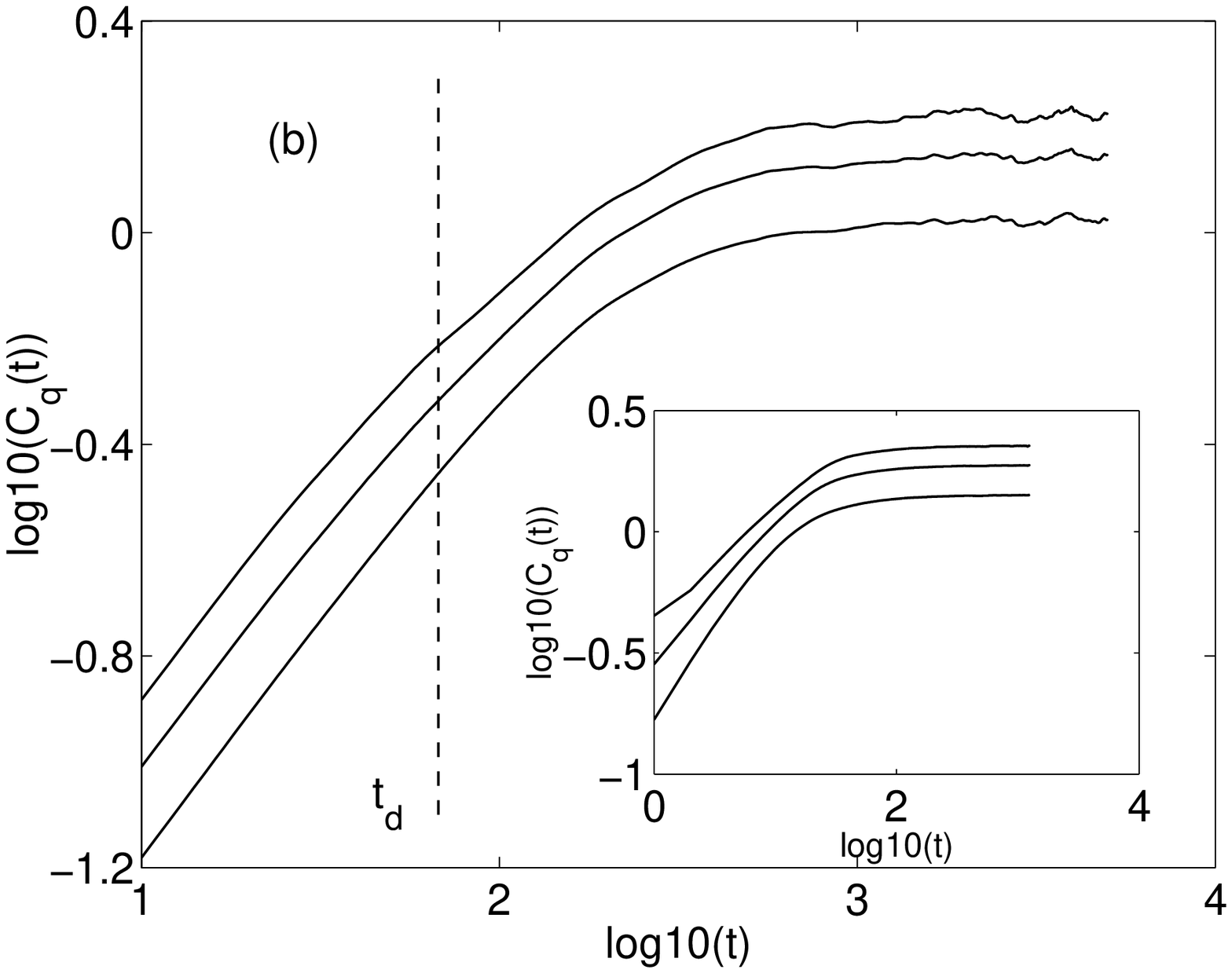}
\caption{(a) The temporal correlation function $C_2(t)$ for $L =
256$, with $\Delta \alpha = 0.2$ and $v=0.015, 0.01$, and $0.005$.
In the inset, the corresponding data are shown for the linearized
interface equations. (b) Temporal correlation functions $C_q(t)$
with $q = 2, 4$, and $6$ for $L = 256$, with $\Delta \alpha = 0.2$
and $v=0.015$. In the inset, the the corresponding functions are
shown for the linearized interface equations. See text for
details.} \label{TemporalFigure}
\end{figure}

The interface fluctuations in the presence of quenched disorder
were analyzed by numerical integrations of Eq. (4) with the
appropriate boundary conditions shown in the schematic
representation of Fig. \ref{Configuration}. In order to implement
the von Neumann boundary condition at $y=0$, Eq. (4) is modified
as follows:
\begin{eqnarray}
\partial_t \phi({\bf x},t) & = &  \nabla^2 \left[
- \phi({\bf x},t) + \phi^3({\bf x},t) - \nabla^2 \phi({\bf x},t) -
F H_0(t)-\alpha({\bf x})\right],
\label{Num}
\end{eqnarray}
where the addition of the term $F H_0(t)$ ensures that the
interface will propagate at constant velocity $v \equiv \dot H_0 =
F/2$ due to the boundary condition $\partial_y \mu|_{y=0} = -F$.
The above equation is then solved with the Dirichlet boundary
condition $\mu|_{y=0} = 0$. This is a numerical trick invoked to
obtain a chemical potential profile consistent with the von
Neumann boundary condition in the domain from the reservoir at
$y=0$ to the interface at $y=H$. The position of the interface
$H(x,t)$ at each $x$ was defined as $H(x,t)=0$ by linear
interpolation between the points of the numerical grid. Without
any loss of generality, the chemical potential at the boundary is
chosen such that $\mu(x,y=0)=0$, leading to $\phi (x,y=0) =
\phi_0$, where $\phi_0$ is the solution of $-\phi_0 + \phi_0^3 = F
H_0(t)$ \footnote{It should be pointed out that the von 
Neumann boundary condition can
also be implemented directly by using $\mu(y = 0) = \mu(y = \Delta
y) + F \Delta y$, where $\Delta y$ is the size of the spatial
discretization. The numerical implementation of the von Neumann
boundary condition is then used to fix $\phi(x, y = 0) = \phi_0$,
where $\phi_0$ is the solution of $-\phi_0 + \phi_0^3 = -\phi(x,y =
\Delta y) + \phi(x,y = \Delta y)^3 + F \Delta y$. We have
compared the two different implementations and found no
distinguishable differences.}.

Typical plots for the chemical potential and the density field
along the $y$ axis obtained from numerical integration of Eq.
\eqref{Num} without quenched disorder ($\Delta \alpha = 0$) are
shown in Figs. \ref{Configuration} (b) and (c), respectively. A
set of successive interface configurations with $v = 0.02$, and
$\Delta \alpha = 0.2$ are also shown in Fig. \ref{Configuration}
(d). It is interesting to note from Fig. \ref{Configuration}(c)
that behind the moving interface the density field has a finite
slope, which remains constant in time. This is due to the
underlying local conservation law for $\phi(x,y)$. We note that
the slighly growing value of $\phi$ which is greater than $+1$ is
just a numerical artifact and can be removed by using a method of
moving box to hold the interface at a constant height, {\it i.e.}
pulling down the disorder field at constant time intervals.
%Also, one could consider a modification
%to the chemical potential that would introduce an asymptotic
%divergence at some $\phi = 1 + \epsilon$, keeping $\phi$ close
%to 1 even as the chemical potential increases in time.

\subsection{Spatial roughness}

With $\Delta \alpha > 0$, the driven wetting front kinetically
roughens as can be seen in Fig. 1(d). To characterize the spatial
extent of the roughness, we first consider the spatial two-point
correlation function
\begin{eqnarray}
G_2(r,t) = \langle \overline{[h(x+r,t) - h(x,t)]^2}
\rangle^{1/2},
\end{eqnarray}
which is directly related to the structure factor $S(k,t) \equiv
\langle \overline{h_k (t)  h_{-k} (t)} \rangle$. In the above
equations the brackets denote an average over different
configurations of random noise, and the overbar a spatial average
over the system.

In Fig. \ref{CorrelationFigure}(a) we show numerical data for the
spatial correlation function. We find that the correlation length
of the roughness of the interface saturates after an initial
growth. According to Eq. \eqref{cross}, the crossover length
$\xi_\times$ is related to the interface velocity by $\xi_\times
\sim v^{-1/2}$. In the inset of Fig. \ref{CorrelationFigure}(a) we
plot the velocity dependence of the corresponding crossover length
$\xi_2$ found from $G_2(r)$. We indeed find that this length
$\xi_2 \sim v^{-0.45}$, which means that $\xi_2 \propto
\xi_\times$ as in the case of spontaneous wetting, too
\cite{dubeprl}.

An estimate for the global roughness exponent $\chi$ can be
obtained from the structure factor $S(k,t)$, as shown in Fig.
\ref{CorrelationFigure}(b). As expected, we find that $S(k) \sim
1/k^{1+2\chi}$ is well satisfied, with a global roughness exponent
of $\chi \approx 1.25 $, and a crossover to a plateau
corresponding to distances larger than the intrinsic correlation
length $\xi_\times$, consistent with
the analysis from the linearized interface equation. We actually
found that the global roughness exponent slightly depends upon the
velocity and increases with decreasing velocity until
asymptotically approaching a value of about $1.35$. For velocities
$v=0.005, 0.002$, and $0.001$, with $\Delta \alpha = 0.1$, 
it was found that  $\chi \approx 1.27,
1.35$, and $1.37$, respectively. Moreover, that global roughness
exponent slightly depends on the strength of the noise. For
example, for $v=0.005$, $\chi \approx 1.27$ and  $1.36$ for
$\Delta \alpha = 0.1$ and $0.2$, respectively. We also compared
the crossover lengths obtained from $G_2(r,t)$ to those from
$S(k,t)$,
%
%The crossover length located
%from $S(k,t)$ more or less agrees with that predicted by
%Eq. ($\ref {cross}$). However, it is difficult to pinpoint the
%exact location of the crossover  from $S(k,t)$
%because there are very few data points in the plateau corresponding
%to large length scales or short wave-vector. On the other hand,
%
and found that the value of $\xi_\times$ from $S(k,t)$ is
about one-half of that from $G_2(r,t)$, independent of disorder strength. 

We also estimated the the local roughness exponent $\chi_{\textrm{loc}}$ 
according to the scaling relationship 
$G_2(r = 1, t) \sim \xi_\times^{\chi-\chi_{\textrm{loc}}} \sim 
v^{(\chi_{\textrm{loc}}-\chi)/2}$ for different velocities \cite{dubeprl}, 
and found that 
$\chi_{\textrm{loc}} \approx 1.0$, as expected for superrough interfaces. 
The spatial 
correlation funnction $G_2(r,t)$ should also follow the same scaling form 
$G_2(r,t) \sim \Delta \alpha v^{-\chi/2}g(rv^{1/2})$ as 
$G_2(r,H)$ for spontaneous imbibition with fixed interface front height 
\cite{dubeprl}. 

%
%This curious(?) fact can also
%be extracted from the linearized interface equation with thermal disorder,
%where the structure factor is analytically solvable and $G_2$ can be
%obtained numerically by a cosine transformation. The factor between
%the crossover lengths is independent of disorder strength, but has
%a value ranging from $3$ to $3.5$. depending on what???
%

\subsection{Temporal roughness}

To quantify the temporal development of the roughness, we consider
the width of the interface defined by
\begin{eqnarray}
w^2(t) =  \langle \overline {(h(x,t)-
\overline {h(x,t)})^2 } \rangle.
\end{eqnarray}
In the presence of quenched disorder the roughness initially
increases as a power law of time, as shown in Fig.
\ref{DataCollapse}. After a crossover, the roughness reaches a
saturated regime. For small $\Delta \alpha = 0.1$, only relatively
low velocities were studied, because for velocity as high as
$v=0.05$, the roughness profile shows pronounced oscillations. The
cause of such numerical oscillations was identified to be the
numerical interpolation of the interface position between the grid
points with time scale equal to lattice size over the velocity. It
can be clearly seen from the inset of Fig. \ref{DataCollapse} that
for the same noise strength, all the roughness curves follow the
same initial growth profile, suggesting an universal growth
exponent $\beta$. It was found that for $\Delta \alpha = 0.1,
0.2$, and $0.3$, the corresponding values of $\beta$ are about
$0.48, 0.50$, and $0.52$, respectively.

The saturated width of the interface was found to be independent of
the lateral system size as long as the lateral system size is larger
than the intrinsic crossover length $\xi_\times$, which has been
derived from the linearized interface equation in the preceding
section. It should be pointed out that in Ref. \cite{alava04}, where
the case of driven wetting was briefly discussed, it is claimed that
after an initial growth, the interface roughness follows a weak
logarithmic growth. From Fig. \ref{DataCollapse} one can see that we
do not find any evidence of such a logarithmic growth regime,
although it could be too slow to be detected numerically.
%It should be noted that we cannot see a clear reason as to why $xi_x$ should
%be a hard upper limit for the lateral length of correlations, as opposed
%to a change in different type of growth regime.

Assuming that the crossover length $\xi_\times$ controls the
roughening process, we can use the results in Ref. \cite{dubeprl}
and write a Family-Vicsek type of scaling relation
\begin{equation}
\label{CollapseForm}
w(t)=\Delta\alpha\xi_\times^{\chi}g\left(\frac{t^{\beta}}{\xi_\times^{\chi}}\right).
\end{equation}
Data collapse using this scaling from is presented in Fig.
\ref{DataCollapse}. Using the data collapse, we give our best
estimates for the roughness and growth exponents as $\chi  = 1.35
\pm 0.05$ and $\beta = 0.50 \pm 0.02$. We note that Ref.
\cite{alava04} estimates that $\chi \approx 1.25$ and $\beta \approx
0.4$, with $\beta=\chi/3$. However, our results do not support this
relation.
%
%A larger, if still slight, discrepancy
%from this linear dependence is found in the interface model itself.
%

In the case of spontaneous imbibition, it was found that the rough
interfaces obey temporal multiscaling, with different scaling
exponents for different moments of the time-dependent correlation
functions \cite{dubeprl}. These functions and the corresponding
exponents are defined by
\begin{eqnarray}
C_q(t) =   \langle \overline{[H(x,t+s) -H_0(t+s)- H(x,s)+H_0(s)]^q}
\rangle^{1/q},
\end{eqnarray}
for $q=1,2,3,...$. The correlation functions $C_2(t)$ are shown in
Fig. \ref{TemporalFigure}(a) for different velocities. The crossover
time $t_\times$ between the power law regime of $C_2(t)$ and
saturation increases with decreasing velocities. The different
functions $C_q(t)$ for $q = 2, 4$ and $6$ are shown in Fig.
\ref{TemporalFigure}(b). At early times all the functions follow
power law behavior $C_q(t) \sim t^{\beta_q}$, with exponents $\beta_q \approx 0.94$
which are independent of $q$. 
This behavior is, however, observed only in time scales
smaller than the disorder persistence time $t_d=\Delta y/v$, where
$\Delta y  = 1$ is the dimensionless spatial discretization step. If
we consider $C_q(t)$ for $t>t_d$, we find evidence multiscaling with
$\beta_2 \approx 0.79$, $\beta_4 \approx 0.69$ and $\beta_6 \approx 0.54$ 
in a small time regime, similar to the spontaneous case
\cite{dubeprl}. However, it is difficult to verify true multiscaling
here because of the rapid crossover to the saturated regime,
although we do expect avalanche type of motion to be present here,
which often leads to multiscaling behavior \cite{leschhorn94}.

\subsection{Numerical Results from the Linearized Interface Equations}
\label{numerics_interface}

An interesting question concerns the range of validity of the
linearized interface equation (LIE), Eq. \eqref{fluct_int}, for the
fluctuations of the height, in particular in the driven, nonlinear
regime with noise included. This issue is also related to the
possible existence of universality classes of roughening for
conserved systems with quenched noise; a problem for which there are
virtually no analytical results. To this end, we have integrated Eq.
\eqref{fluct_int} numerically in time. To incorporate the nonlinear
nature of the disorder $\eta(x,H(x,t))$, a back-and-forth Fourier
transformation scheme is required at every time step. A square
lattice landscape of independently Gaussian distributed noise was
used for $\eta(x,y)$. We note that solving Eq. \eqref{fluct_int}
instead of the full 2D phase field model is numerically much easier.

To compare the results with the phase-field model, we numerically
computed the same set of correlation functions. From the analytic
derivation of the interface equation, we can obtain a quantitative
map between the parameters of the phase field model and the
interface model, which is as follows. The interface velocities
should obviously be the same. The disorder fields between the models
are related by $\eta=M\alpha$, where $M$ is the mobility in Fick's
law. In our dimensionless units $M=1$. The effective surface tension
in the phase field model is given by the standard form of the
potential $V(\phi)$ as $\sigma_{pf}=\int
du~(\partial_u\phi_0(u))^2$, where $\phi_0$ is the 1D kink solution
of the disorder-free system. The surface tension in the interface
equation comes out as $\sigma=M\sigma_{pf}$.

The results from the LIE are collected in the insets of the
corresponding phase field results in Figs. 2-4. Obviously, the
length scale $\xi_\times$ is present in an identical manner in both
cases. Remarkably enough, the corresponding results from the two
cases are in most cases quantitatively close to each other. There
are some important differences, however. First, the saturated
interface widths differ by about $20\%$, the interfaces from the LIE
being rougher. Since the interface model only takes into account
linear dissipation effects, we would expect it to underestimate the
stiffness of the interface, leading to larger roughness amplitudes
than in the full phase field model.

Next, we examine the data collapse for the interface roughness (Fig.
\ref{DataCollapse}) using the scaling function of
Eq.~\eqref{CollapseForm}. The best collapse is obtained for slightly
different values of the exponents as compared to the phase-field
model. The roughness exponent is close to the previous value, namely
$\chi=1.27 \pm 0.05$. However, the growth exponent is now given by
$\beta=0.37 \pm 0.04$ for the LIE, where the difference to the phase
field result is outside of the numerical uncertainties. In the LIE
we also noted a slight deviation from the behavior of the crossover
time $t_\times \propto \xi_\times^{\chi/\beta}$ as a function of the
disorder strength $\Delta\alpha$.
% ???
%as well as a $\Delta\alpha$
%exponent of $1.08\pm0.03$ instead of $1$.
%
Another difference between the two models was found in temporal
multiscaling. In the LIE each of the moments increases with a
different exponent $\beta_q$ even at times smaller than the disorder
persistence length $t_d$. The exponents are $\beta_2 = 0.75
\pm0.03$, $\beta_4 = 0.55 \pm 0.05$, and $\beta_6 = 0.47 \pm 0.03$.
%This multiscaling has been linked to avalanche type motion of the
%interface \cite{Leschhorn94}, which should be present in our system.
The values of these exponents are lower than in the phase-field
model, which is not surprising since the LIE has a lower value of
$\beta$, too.

\section{Summary and Conclusions}
\label{conclusions}

In the present work, we have studied wetting of a disordered medium
driven by a constant mass flux in a 2D system. Our model is a
prototype phase field model incorporating mass conservation into the
flow of two immiscible fluids, Eq. \eqref{pf_eq}. From this model we
have derived non-local interface equations to lowest order in
Fourier space fluctuations, Eqs. \eqref{ave_int} and
\eqref{fluct_int}. Because of the linearization, these equations are
local in Fourier space. The constant flux boundary condition gives
rise to a interface that moves with a constant velocity proportional
to the flux. We have obtained a time-independent crossover length
scale $\xi_\times \propto \sqrt{\sigma/v}$ from the
interface equations. Numerically, we find that the kinetic
roughening of an interface is governed by a scaling relation of the
Family-Vicsek type, where $\xi_\times$ controls the extent of the
fluctuations (for $\xi_\times < L$) as given by Eq.
\eqref{CollapseForm}. For the kinetic roughening of the interfaces,
we find that they are superrough, with $\chi = 1.35 \pm 0.05$ and
$\chi_{\rm loc} \approx 1$. For
temporal roughness, $\beta=0.50 \pm 0.02$, and there is numerical
evidence of temporal multiscaling for $t>t_d$. We note that all
these results are in qualitative but {\it not in quantitative}
agreement with the case of spontaneous imbibition studied earlier in
Refs. \cite{dubeprl} for interfaces, which are kept at
constant height in the steady-state regime described by Washburn's
law.

In addition to obtaining the LIE by analytic methods, we have also
made a direct comparison between it and the full phase field model
with quenched noise properly included. We find very good agreement
between the spatial correlations of the interfaces, even including
approximately the same roughness exponent of $\chi \approx 1.3$.
However, the temporal correlations in the two cases are different:
while the amplitude of the saturated roughness is larger in the LIE,
but the growth exponent $\beta = 0.37 \pm 0.04$ is smaller than the
phase-field result $\beta = 0.50 \pm 0.02$ in the phase field model.
Also, spatial multiscaling is more clearly present in the LIE within
our numerics.

Our analytical and numerical results of forced wetting can be
compared with experimental results of kinetic roughening of an
oil-air interface in a forced wetting where the experiments were
done in a horizontal Hele-Shaw cell with quenched disorder
\cite{SOH1}. It was found in the experiments that the growth
exponent $\beta \approx 0.5$ which is nearly independent of the
experimental parameters, and the roughness exponent $\chi \approx
1.3$, which, however, depends on experimental parameters. While a
fully quantitative comparison may be difficult, both exponents
obtained experimentally are in very good agreement with our
numerical results. Furthermore, the experiments confirm that the
crossover length scales as the inverse of the square root of
velocity, as found in our theory. Further experiments on the
dependence of the results on other systems parameters would be most
interesting.

\section*{Acknowledgments}
This work has been supported in part by the Academy of Finland
through its Center of Excellence grant. We would like to thank M.
Alava and M. Dube for their insightful comments.

\section*{Appendix}
\appendix

\section{Projection to an Interface Equation}
\label{A}

In this Appendix we discuss the details of the sharp interface
projection to obtain the interface equation, Eq. \eqref{non-local}
from the phase field equation of motion, Eq. \eqref{pf_eq}. First we
invert the phase field equation in a volume $V$ with boundary $S$,
by multiplying with the Green's function, integrating over $V$ and
applying Gauss's divergence theorem. The result is
\begin{equation}
\label{general pf} \int_V d^3r'\sqrt{\det(g')}
G(r,r')\partial_t\phi(r')=\mu(r)+\Lambda,
\end{equation}
where the surface term $\Lambda$ is explicitly given by
\begin{equation}
\Lambda=\int d\vec{S'}\cdot
[G(r,r')\nabla'\mu(r')-\mu(r')\nabla'G(r,r')].
\end{equation}
The standard procedure is then to consider a single-valued sharp
interface $H(x,t)$, and transform to coordinates of distance along
and perpendicular to this interface given by $(s,u)$. In these
coordinates the metric tensor is given by
\begin{equation}
g=\begin{bmatrix}
1 & 0 \\
0 & (1+u\kappa)^2
\end{bmatrix},
\end{equation}
where $\kappa$ is the curvature of the interface, defined via the
unit tangent $\mathbf{t}$ and unit normal $\mathbf{n}$ of the
interface as $\kappa\mathbf{t}=\partial_s\mathbf{n}$. The volume
integration measure is the Jacobian $J=\sqrt{\det(g)}=1+u\kappa$,
and it must be positive definite. This limits the validity of the
coordinates to the area not further from the interface than the
radius of the interface curvature.

Next, a number of standard approximations are made, including the
small curvature approximation, which gives the Laplacian to first
order in $\kappa$ as
\begin{equation}
\nabla^2\simeq \frac{\partial^2}{\partial u^2}+\frac{\partial^2}{\partial s^2}+
\kappa\frac{\partial}{\partial u}.
\end{equation}
For a sharp interface with small curvature, the phase field near the
interface has the form given by the $\alpha=0$, 1D kink solution
$\phi_0$ in the normal direction, defined by
$\partial_u^2\phi_0=V'(\phi_0)$. The chemical potential is then
$\mu\simeq -\kappa\partial_u\phi_0-\alpha$. Since the kink solution
has a small gradient except at the interface, we can project Eq.
\eqref{general pf} to the interface with the operator $\int
du\partial_u\phi_0(u)[\cdot]$, and take the explicit sharp interface
limit $\phi_0\simeq-1+2\Theta(u)$. The projected equation involves
contributions only from an area not further from the interface than
the interface width, which is less than the interface radius of
curvature by the virtue of the small curvature and sharp interface
approximations. Therefore, the use of coordinates $(s,u)$ is valid.
Eq. \eqref{general pf} is then projected to
\begin{equation}
\label{int_int}
2\int ds'G(s,0\vert s',0)\partial_tu(s')=-\sigma\kappa-\alpha(s,0)+
\Lambda\vert_{u=0},
\end{equation}
where $\sigma=\frac{1}{2}\int du~(\partial_u\phi_0(u))^2$ is the
effective surface tension. This can be transformed to Cartesian
coordinates using $ds\partial_tu=dx\partial_tH(x,t)$, yielding Eq.
\eqref{non-local}.

\section{Linearization of the Interface Equation}
\label{B}

In this Appendix we describe in detail the method of using strip
geometry to obtain the linearized interface equations (Eqns.
\eqref{ave_int} and \eqref{fluct_int}) from the full non-local sharp
interface equation, Eq. \eqref{non-local}. For the half-strip
$\{(x,y)\vert x\in [0,L],y\in [0,\infty]\}$, the Green's function
for the Laplacian, with homogenous von Neumann boundary conditions
at the strip edges, is given by
\begin{equation}
\begin{split}
\label{SGreen}
G(x,y\vert x',y')=\frac{1}{2L}\left[\vert y-y'\vert+y+y'\right]\\
-\frac{1}{\pi}\sum_n\frac{1}{n}\cos\left(\frac{n\pi x}{L}\right)
\cos\left(\frac{n\pi x'}{L}\right)
\left[e^{-\frac{n\pi}{L}\vert y-y'\vert}+e^{-\frac{n\pi}{L}(y+y')}\right].
\end{split}
\end{equation}
The boundary term is readily evaluated as $\Lambda=F\int dx'
G(x,y;x',0)=Fy$. The linearization can only be done around the
interface of the disorder-free system. This is the same as the
average interface height of the disordered system only when $F$ is
much larger than the critical driving force of the underlying
pinning-depinning transition. For a disorder-free system Eq.
\eqref{non-local} becomes
\begin{eqnarray}
2\int dx' G(x,H_0(t);x',H_0(t))\partial_tH_0(t)&=&FH_0(t)\\
\Leftrightarrow\qquad\partial_tH_0(t)&=&\frac{MF}{2}\label{pure}.
\end{eqnarray}
Linearizing Eq. \eqref{non-local} using $H(x,t)=H_0(t)+h(x,t)$ leads
to
\begin{eqnarray}
I_B+I_C+I_D&=&-\frac{\sigma}{2}\partial_x^2h(x,t)-\frac{1}{2}\eta (x,H(x,t))+
\frac{1}{2}MFh(x,t); \label{Sfluct}\\
I_B&=&\int dx' \partial_yG(x,y;x',H_0)\vert_{y=H_0}h(x,t)\partial_tH_0;\\
I_C&=&\int dx' \partial_{y'}G(x,H_0;x',y')\vert_{y'=H_0}h(x',t)\partial_tH_0;\\
I_D&=&\int dx' G(x,H_0;x',H_0)\partial_th(x',t).
\end{eqnarray}
The derivatives of $G$, as obtained from the definition of Eq.
\eqref{SGreen}, are discontinuous at $y=y'=H_0(t)$, where the
linearization was done. To go around this problem we simply set
$\Theta(y-y')\vert_{y=y'=H_0(t)}=1/2$.
%The correct form of the
%linearization would take into account on which side of the
%discontinuity the linearization is done, i.e. linearize in two
%different branches and choose the correct one according to $h(x,t)$.
%This produces factors  $\Theta(h(x,t))$, and $\Theta(h(x',t))$, to
%$\partial_yG$ and $\partial_{y'}G$, respectively. Since these lead
%to complicated $h$ dependence, we make the rather crude
%simplification.
We have also performed the same half-strip linearization in the case
of the spontaneous imbibition, where the half-plane Green's function
can also be linearized directly \cite{dubeprl,dube00a}. These two
methods yield identical results, {\it i.e.} linearization and
Fourier transformation commute with the half-plane limit.

The rest of the procedure is then straightforward, and by defining
the Fourier series representations
\begin{eqnarray}
h(p,t)&=&\frac{1}{L}\int dx \cos\left(\frac{p\pi x}{L}\right)h(x,t);\\
\eta(p,t)&=&\frac{1}{L}\int dx \cos\left(\frac{p\pi x}{L}\right)\eta(x,H(x,t)),
\end{eqnarray}
and projecting Eq.~\eqref{Sfluct} to Fourier component $p$ with
$P_p[\cdot]=\int dx\cos\left({p\pi x}/{L}\right)[\cdot]$, we obtain
\begin{equation}
\begin{split}
\partial_tH_0h(p,t)\left[1+e^{-2\frac{p\pi}{L}H_0}\right]-\frac{L}{p\pi}\partial_th(p,t)
\left[1+e^{-2\frac{p\pi}{L}H_0}\right]=\\\sigma\left(\frac{p\pi}{L}\right)^2 h(p,t)-\eta(p,t)+Fh(p,t).
\end{split}
\end{equation}
In the Fourier projection of the curvature term, one obtains
boundary terms that are non-zero at non-zero contact angles, but
they are negligible in the limit $L\rightarrow\infty$. Changing
variables to $k={p\pi}/{L}$ and substituting $F=2\partial_tH_0$ we
obtain Eq. \eqref{fluct_int}, with a discrete wave vector $k$.
Taking the limit $L\rightarrow\infty$ while keeping $k$ constant
finally gives the proper continuum limit.

%\bibliographystyle{apsrev}
%\bibliography{references}

%\begin{comment}

%\end{comment}

%\newpage

\end{document}